\documentclass[%
reprint,
superscriptaddress,
nofootinbib,
amsmath,amssymb,
]{revtex4-1}

\pdfoutput=1
\usepackage{float}
\usepackage{xcolor}
\usepackage{graphicx}
\usepackage{dcolumn}
\usepackage{bm}
\usepackage[colorlinks=true, linkcolor=blue, citecolor=blue, urlcolor=blue]{hyperref}
\newcommand{\ZIB}{Zuse Institute Berlin, 14195 Berlin, Germany}
\newcommand{\JCM}{JCMwave GmbH, 14050 Berlin, Germany}

\usepackage{graphicx}
\usepackage{booktabs}
\usepackage{tabularx}

\newcommand{\mytoprule}{\specialrule{0.1em}{0.2em}{0.2em}}
\newcommand{\mymidrule}{\specialrule{0.1em}{0.2em}{0.2em}}
\newcommand{\mybottomrule}{\specialrule{0.1em}{0.2em}{0.2em}}

\begin{document}

\title{Computation of eigenfrequency sensitivities using Riesz projections
for\\ efficient optimization of nanophotonic resonators}

\author{Felix Binkowski}
\affiliation{\ZIB}
\author{Fridtjof Betz}
\affiliation{\ZIB}
\author{Martin Hammerschmidt}
\affiliation{\JCM}
\author{Philipp-Immanuel Schneider}
\affiliation{\JCM}
\author{Lin Zschiedrich}
\affiliation{\JCM}
\author{Sven~Burger}
\affiliation{\ZIB}
\affiliation{\JCM}

\begin{abstract}
\vspace{0.15cm}
Resonances are omnipresent in physics and essential for the description of wave phenomena.
We present an approach for computing eigenfrequency sensitivities of resonances.
The theory is based on Riesz projections and the approach can be applied to
compute partial derivatives of the complex eigenfrequencies of any resonance problem.
Here, the method is derived for Maxwell's equations.
Its numerical realization
essentially relies on direct differentiation of scattering problems.
We use a numerical implementation to demonstrate the performance of the approach compared
to differentiation using finite differences. 
The method is applied for the efficient optimization of the quality factor of a nanophotonic resonator.
\vspace{0.8cm}
\end{abstract}

\maketitle

\section{Introduction}
\label{sec:intro}

Resonance phenomena are ubiquitous in nanophotonics
and play an important role for tailoring light-matter
interactions~\cite{Novotny_NatPhot_2011,Kuznetsov_DielectricNanostruc_2016}.
They are exploited in, e.g.,
single-photon sources for quantum technology~\cite{Senellart_2017},
biosensors~\cite{Anker_BioSens_NatMater_2008},
nanolasers~\cite{Ma_NatNanotechnol_2019},
or solar energy devices~\cite{Ma_LiSciAppl_2016,Zhang_ChemRev_2018}.
All these applications rely on the highly localized electromagnetic field energies in the vicinity of
the underlying nanoresonators~\cite{Lalanne_QNMReview_2018}. A central figure of merit
for the description of resonance effects is the quality ($Q$) factor,
which quantifies, in the case of low-loss systems, the relation between stored and radiated field energies
of the resonances~\cite{Wu_ACSPhot_2021}.
Nanoresonators with low energy dissipation, i.e., with high $Q$-factors, have been proposed to improve
the efficiencies of nanophotonic devices~\cite{West_LasPhotRev_2010,Kuznetsov_DielectricNanostruc_2016}.
For example, high-$Q$ resonators can boost the brightness
of quantum emitters, the sensitivity of sensors, or the emission processes in plasmonic lasers~\cite{Wang_highQ_2021}.
Designing devices with numerical optimization is a time and cost effective approach.
The resonances are numerically computed by solving the source-free Maxwell's equations equipped
with open boundary conditions~\cite{Lalanne_QNM_Benchmark_2018}. This yields non-Hermitian eigenproblems and
the solutions are eigenmodes with complex-valued eigenfrequencies.
In this context, the $Q$-factor
is defined as the scaled ratio of the real and imaginary parts of the eigenfrequency.
\footnote{This work has been published:\\
F. Binkowski et al., Commun. Phys. \textbf{5}, 202 (2022).\\
DOI: \href{https://doi.org/10.1038/s42005-022-00977-1}{10.1038/s42005-022-00977-1}}

Nanoresonators with high $Q$-factors have been theoretically presented,
but fabrication of these resonators is a limiting task~\cite{Wang_highQ_2021}. 
The sensitivity analysis of eigenfrequencies can show a way to reduce the sensitivities
of the $Q$-factors. This can support the nanofabrication processes.
Furthermore, the sensitivity analysis of eigenfrequencies is essential for numerical simulation.
For example, the numerical accuracies of the calculated eigenfrequencies are strongly
influenced by the sensitivities of the eigenfrequencies when the systems are subject
to small perturbations~\cite{Bindel_2013,Guettel_NLEVP_2017}.
In particular, for high-$Q$ resonators,
the accuracy requirements are demanding since the real
and imaginary parts of the eigenfrequencies differ by several orders of magnitude.
Sensitivities are also directly exploited in numerical
optimization algorithms using gradients~\cite{Jensen_LasPhotRev_2011},
for gradient-enhanced surrogate modelling~\cite{Bouhlel_2019},
and for local sensitivity analyses~\cite{Cacuci_2005}.
The computation of eigenfrequency sensitivities is usually based
on perturbation theory~\cite{Kato_1995,Sakurai_Napolitano_2020},
where the sensitivity
of the underlying operator, the left and the right eigenmodes,
and a proper normalization of the eigenmodes are required.
The solution of the perturbed systems, on the other hand, is not necessary.
For resonance problems,
left and right eigenmodes are in general not
identical, which increases the computational effort, and normalization requires
additional attention. 
There are specialized approaches that,
e.g., exploit magnetic fields for extracting the left eigenmodes~\cite{Burschaepers_2011},
introduce an adjoint system for computing sensitivities~\cite{Swillam_2008},
or that rely on internal and external electric fields at the boundaries of the nanoresonators~\cite{Yan_2020}.
It is also possible to completely omit the use of eigenmodes for sensitivity analysis~\cite{Alam_2019}.
A further approach is the straightforward application of finite differences.
However, this also includes the solution of the perturbed resonance problems,
which increases the computational effort.

\begin{figure*}[]
\includegraphics[width=0.95\textwidth]{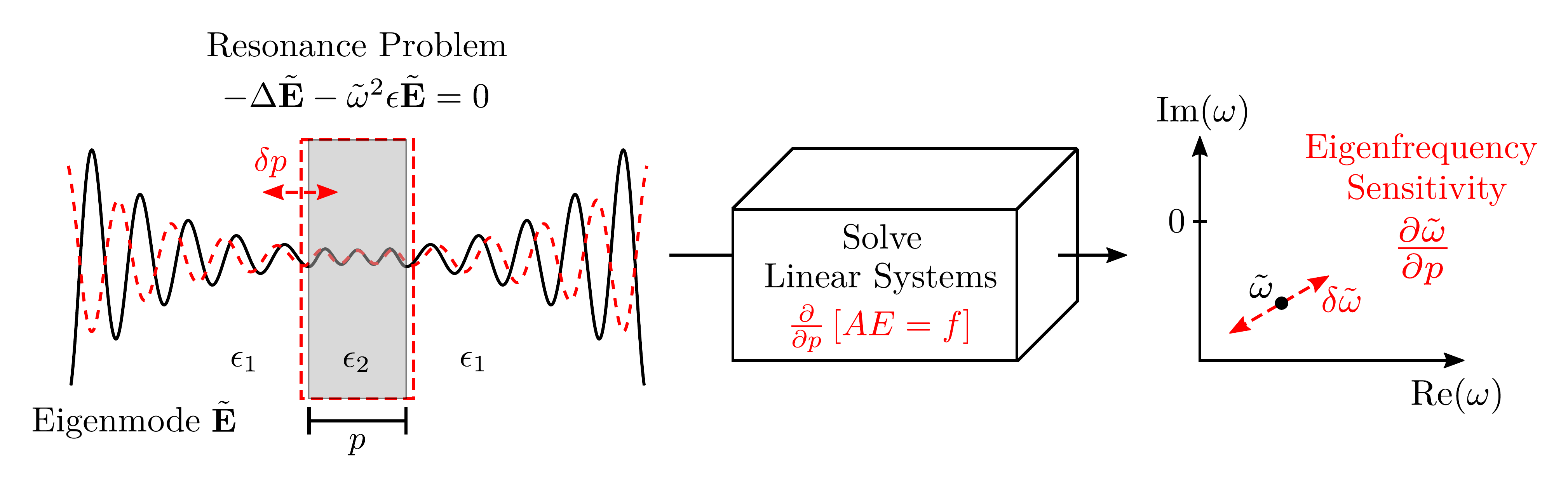}
\caption{\label{fig01} 
Schematic representation of computing eigenfrequency sensitivities of
a resonator using contour integration.
The system is defined by layers with different
permittivities $\epsilon_1$ and $\epsilon_2$ and is described by 
the one-dimensional Helmholtz equation
$-\Delta\tilde{\mathbf{E}}-\tilde{\omega}^2\epsilon\tilde{\mathbf{E}}=0$.
A solution to the resonance problem
is given by the eigenmode $\tilde{\mathbf{E}}$
and the corresponding complex-valued eigenfrequency $\tilde{\omega}\in \mathbb{C}$.
The real part of the electric field of the eigenmode is sketched with the solid black curve.
A perturbation $\delta p$ of the middle layer width $p$ leads
to a perturbed electric field, represented by the dashed red curve, and to 
a perturbation $\delta \tilde{\omega}$ of the eigenfrequency.
Computing contour integrals by solving linear systems $AE=f$ and $\partial / \partial p \left[AE=f\right]$
in the complex frequency plane
yields the eigenfrequency sensitivity $\partial \tilde{\omega}/\partial p$.
Solving the linear systems is considered as a blackbox.}
\end{figure*}

In this work, we present an approach for computing eigenfrequency sensitivities
that completely avoids solving resonance problems.
The approach is based on Riesz projections
given by contour integrals in the complex frequency plane. The contour
integrals are numerically accessed by solving Maxwell's equations with a source term
enabling an efficient numerical realization using direct differentiation.
The numerical
experiments show a significant reduction in computational effort compared to applying finite differences.
A Bayesian optimization algorithm with the incorporation of
eigenfrequency sensitivities is used to optimize a resonator hosting a resonance with a high \mbox{$Q$-factor}.

\section{Theoretical background and numerical realization}
\label{sec2}
We start with an introduction of the theoretical background on
resonance phenomena occurring in nanophotonics. Based on this,
Riesz projections for computing eigenfrequency sensitivities and
an efficient approach for its numerical realization are presented.

\subsection{Resonances in nanophotonics}
In nanophotonics, in the steady-state regime, light-matter interactions can be described
by the time-harmonic Maxwell's equations in second-order form,
\begin{align}
	\nabla \hspace{-0.05cm} \times \hspace{-0.05cm} \mu_0^{-1} 
	\nabla \hspace{-0.05cm} \times \hspace{-0.05cm} \mathbf{E}(\mathbf{r},\omega_0) \hspace{-0.05cm} - \hspace{-0.05cm}
	\omega_0^2\epsilon(\mathbf{r},\omega_0) \mathbf{E}(\mathbf{r},\omega_0) \hspace{-0.05cm} = \hspace{-0.05cm}
	i\omega_0\mathbf{J}(\mathbf{r}), \label{maxwell}
\end{align}
where $\mathbf{E}(\mathbf{r},\omega_0) \in \mathbb{C}^3$ is the electric field,
$\mathbf{r}\in\mathbb{R}^3$ is the position,
$\omega_0\in\mathbb{R}$ is the angular frequency,
and $\mathbf{J}(\mathbf{r})\in \mathbb{C}^3$ is 
the electric current density corresponding to a light source. 
In the optical regime, the permeability tensor $\mu(\mathbf{r},\omega_0)$ typically
equals the vacuum permeability $\mu_0$.
The permittivity tensor 
$\epsilon(\mathbf{r},\omega_0) = \epsilon_\mathrm{r}(\mathbf{r},\omega_0)
\epsilon_0$, where $\epsilon_\mathrm{r}(\mathbf{r},\omega_0)$
is the relative permittivity and $\epsilon_0$
the vacuum permittivity, describes the spatial
distribution of material and the material dispersion.
Solutions to Eq.~\eqref{maxwell} are called scattering solutions as light from a source
is scattered by a material system.

Resonances are solutions to Eq.~\eqref{maxwell}
without a source term, i.e., $\mathbf{J}(\mathbf{r}) = 0$,
and with transparent boundary conditions.
The boundary conditions lead to non-Hermitian eigenproblems, and,
if material dispersion is also present,
the eigenproblems become nonlinear.
The electric field distribution of an eigenmode is 
denoted by $\tilde{\mathbf{E}}(\mathbf{r}) \in \mathbb{C}^3$ and the 
corresponding complex-valued eigenfrequency
by $\tilde{\omega} \in \mathbb{C}$.
The $Q$-factor of a resonance is defined by
\begin{align}
    Q = \frac{\mathrm{Re}(\tilde{\omega})}{-2\mathrm{Im}(\tilde{\omega})} \nonumber
\end{align}
and describes its spectral confinement. In the limiting case of vanishing losses,
this definition agrees with the energy definition, according to which the $Q$-factor quantifies the relation
between stored and dissipated electromagnetic field energy of a resonance~\cite{Wu_ACSPhot_2021}.

In the following, a nanophotonic resonator supporting a resonance with a high $Q$-factor is investigated.
We compute the eigenfrequency sensitivities with respect to various parameters
to optimize the $Q$-factor of the nanoresonator.
Figure~\ref{fig01} sketches the applied framework for
an exemplary problem, a one-dimensional resonator defined by layers with
different permittivities.
Changes $\delta p$ of the parameter $p$
leads to changes in the eigenmode $\tilde{\mathbf{E}}$ and in the corresponding eigenfrequency $\tilde{\omega}$,
which describes the sensitivity of $\tilde{\mathbf{E}}$ and $\tilde{\omega}$ with respect to the
parameter $p$.
To compute the eigenfrequency sensitivity, we introduce a
contour-integral-based approach using Riesz projections,
where physical observables are extracted from scattering problems.
Solving the scattering problems, which are linear systems,
can be regarded as a blackbox~\cite{Binkowski_JCOMP_2020,Betz_2021}.

\subsection{Riesz projections for eigenfrequency sensitivities}
To derive a Riesz-projection-based approach for computing eigenfrequency sensitivities, which
are the partial derivatives of the eigenfrequency,
we consider the electric field $\mathbf{E}(\mathbf{r}, \omega_0 \in \mathbb{R})$
as a solution of 
Eq.~\eqref{maxwell} and $\mathbf{E}(\mathbf{r}, \omega \in \mathbb{C})$
as an analytical continuation of $\mathbf{E}(\mathbf{r}, \omega_0)$
into the complex frequency plane. The field $\mathbf{E}(\mathbf{r}, \omega)$
is a meromorphic function with resonance poles at the eigenfrequencies.
To simplify the notation, we omit the spatial and frequency dependency of
the electric field and write $\mathbf{E}$ when we mean $\mathbf{E}(\mathbf{r}, \omega)$.

Let $\mathcal{L}(\mathbf{E})$ be a physical observable,
where $\mathcal{L}:\mathbb{C}^3 \rightarrow \mathbb{C}$ is
a linear functional, and $\tilde{C}$ be a contour enclosing the 
pole $\tilde{\omega}$ of the order $m$ and no other poles.
Then, the Laurent expansion of $\mathcal{L}(\mathbf{E})$ about $\tilde{\omega}$ is given by
\begin{align}
\begin{split}
    \mathcal{L}(\mathbf{E}) &= \sum_{k=-m}^{\infty}
    a_k(\omega-\tilde{\omega})^k, \hspace{0.5cm} \mathrm{where} \\
    a_k(\tilde{\omega}) &= \frac{1}{2 \pi i}
    \oint \limits_{\tilde{C}} \frac{\mathcal{L}(\mathbf{E}(\omega))}{(\omega-\tilde{\omega})^{k+1}}
    \, d\omega \in \mathbb{C}. 
\end{split}\label{eq:laurent}
\end{align}
The coefficient $a_{-1}(\tilde{\omega})$ is the so-called residue
of $\mathcal{L}(\mathbf{E})$ at $\tilde{\omega}$.
Using Eq.~\eqref{eq:laurent}
with the assumption that $\tilde{\omega}$ has the order $m=1$
and applying Cauchy's integral formula
yield
\begin{align}
    \oint \limits_{\tilde{C}} \omega \mathcal{L}(\mathbf{E})\, d\omega 
    = \oint \limits_{\tilde{C}} \frac{\omega}{\omega-\tilde{\omega}} a_{-1}(\tilde{\omega}) \, d\omega
    = \tilde{\omega}  \oint \limits_{\tilde{C}} \mathcal{L}(\mathbf{E})\, d\omega, \nonumber
\end{align}
where, due to the closed integral in the complex plane, the regular terms in the expansion vanish.
With this, the eigenfrequency $\tilde{\omega}$ is given by
\begin{align}
    \tilde{\omega} = \frac{\oint \limits_{\tilde{C}}\omega \mathcal{L}(\mathbf{E})
    \, d\omega}{\oint \limits_{\tilde{C}} \mathcal{L}(\mathbf{E})
    \, d\omega}. \label{eq:eigenfrequency}
\end{align}
The contour integrals in this equation are essentially Riesz projections for
$\mathcal{L}(\mathbf{E})$ and $\tilde{C}$~\cite{Binkowski_JCOMP_2020}.
Partial differentiation with respect to a parameter $p$ directly gives
the desired expression for the
eigenfrequency sensitivity,
\begin{align}
\begin{split}
    \frac{\partial \tilde{\omega}} {\partial p} =
    \left(\frac{\partial u}{\partial p}\right. v & 
    - u \left.\frac{\partial v}{\partial p}\right) \frac{1}{v^2}, \hspace{0.5cm} \mathrm{where}  \\
    u = \oint \limits_{\tilde{C}}\omega \mathcal{L}(\mathbf{E})
    \, &d\omega, \hspace{0.5cm}
    v = \oint \limits_{\tilde{C}} \mathcal{L}(\mathbf{E})
    \, d\omega,  \\
    \frac{\partial u}{\partial p} =
    \oint \limits_{\tilde{C}}\omega \mathcal{L}\left(\frac{\partial\mathbf{E}}{\partial p}\right)\, &d\omega, \hspace{0.5cm}
    \frac{\partial v}{\partial p} =
    \oint \limits_{\tilde{C}} \mathcal{L}\left(\frac{\partial\mathbf{E}}{\partial p}\right)\, d\omega. 
\end{split}\label{eq:eig_derivative}
\end{align}
For the interchangeability of integral and derivative, $\mathbf{E}$ and
$\partial \mathbf{E}/\partial p$ are assumed to be continuously
differentiable with respect to the frequency $\omega$ and the parameter $p$.
The eigenmode $\tilde{\mathbf{E}}$ and its sensitivity $\partial \tilde{\mathbf{E}}/ \partial p$
can be represented by the contour integrals
\begin{align}
     \tilde{\mathbf{E}} = \oint \limits_{\tilde{C}} \mathbf{E} \, d\omega  \hspace{0.5cm}\mathrm{and}\hspace{0.5cm}
     \frac{\partial\tilde{\mathbf{E}}}{\partial p} =
     \oint \limits_{\tilde{C}} \frac{\partial \mathbf{E}}{\partial p} \, d\omega, \nonumber
\end{align}
respectively, which are Riesz projections applied to Maxwell's equations given by Eq.~\eqref{maxwell}.
This approach can be generalized for multiple eigenfrequencies inside a contour
as well as for higher order poles; $\mathrm{cf.}$ Ref.~\cite{Binkowski_JCOMP_2020}.
Note that Riesz projections can also be used to compute modal expansions of
physical observables, where scattering solutions are expanded into weighted sums of eigenmodes~\cite{Zschiedrich_PRA_2018}.

\subsection{Numerical realization and direct differentiation}
For the numerical realization of the presented approach, the finite element method (FEM) is applied.
Scattering problems are solved by applying the solver \textsc{JCMsuite} \cite{Pomplum_NanoopticFEM_2007}.
The FEM discretization of Eq.~\eqref{maxwell} leads to the linear system of equations
$AE =f$, where $A \in \mathbb{C}^{n \times n}$
is the system matrix, $E \in \mathbb{C}^{n}$ is the scattered electric field
in a finite-dimensional FEM basis, and $f\in \mathbb{C}^{n}$ contains the source term.
The solver employs adaptive meshing and higher order polynomial ansatz functions.
In all subsequent simulations, it is ensured that sufficient accuracies are achieved
with respect to the FEM discretization parameters.
Note that also other methods can be used for numerical discretization. In the field of nanophotonics, 
common approaches are, e.g., the finite-difference time-domain method, the Fourier modal method, 
or the boundary element method~\cite{Lalanne_QNM_Benchmark_2018, Hohenester_2012}.

In order to calculate eigenfrequencies $\tilde{\omega}$ and their 
sensitivities $\partial \tilde{\omega}/\partial p_i$ with respect to parameters $p_i$,
the electric fields $\mathbf{E}$ and their sensitivities $\partial \mathbf{E}/ \partial p_i$
are computed for complex
frequencies $\omega \in \mathbb{C}$ on the contours given in Eq.~\eqref{eq:eigenfrequency}
and Eq.~\eqref{eq:eig_derivative}.
For the calculation of $\partial \mathbf{E}/ \partial p_i$, we apply an approach 
based on directly using the FEM system matrix~\cite{Nikolova_2004,Burger_Derivatives_PROCSPIE_2013}.
With this direct differentiation method, the sensitivities of scattering solutions can be computed by
\begin{align}
\frac{\partial E}{\partial p_i} = A^{-1} \left( \frac{\partial f}{\partial p_i}
- \frac{\partial A}{\partial p_i}E \right). \label{eq:scatt_derivatives}
\end{align}
In a first step, instead of directly computing $A^{-1}$,
an $LU$-decomposition of $A$, which can be seen as the matrix variant
of Gaussian elimination, is computed to efficiently solve the linear system $AE=f$.
In the FEM context, this step is usually a computationally expensive step
in solving scattering problems,
so reusing an $LU$-decomposition can significantly reduce computational costs.
In a second step, 
the partial derivatives of the system matrix, $\partial A/ \partial p_i$, and of the source term,
$\partial f/ \partial p_i$, are obtained quasi analytically, i.e., with negligible computational effort. 
Then, $A=LU$, $E$, $\partial A/ \partial p_i$, and $\partial f/ \partial p_i$ are
used to compute $\partial E/ \partial p_i$ in Eq.~\eqref{eq:scatt_derivatives}.
The $LU$-decomposition can be used to obtain both $E$ and ${\partial E}/{\partial p_i}$.

\begin{figure*}[]
\includegraphics[width=0.95\textwidth]{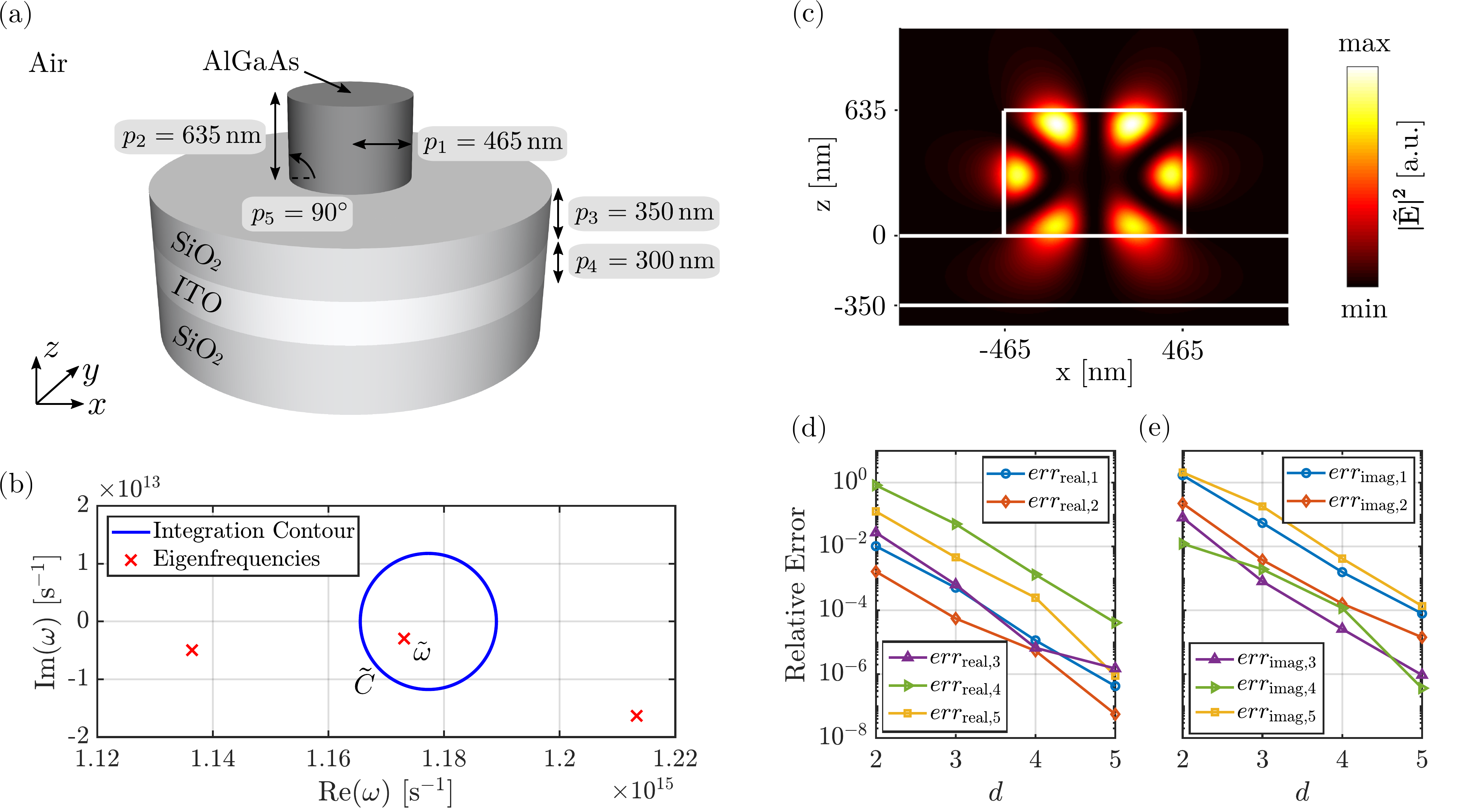}
\caption{\label{fig02} 
Numerical investigation of the high-$Q$ resonance of a nanophotonic resonator.
(a) Nanoresonator on a three-layer substrate.
The substrate is infinitely extended in $x$ and $y$ direction.
The geometrical parameters $p_1,p_2,\dots,p_5$ are the 
reference values from Ref.~\cite{Koshelev_Science_2020}.
(b) Calculated eigenfrequency $\tilde{\omega} = (1.17309 - 0.00296 i)\times 10^{15}\,\mathrm{s}^{-1}$
corresponding to the high-$Q$ resonance. The other red crosses shown  are the two eigenfrequencies
which are closest to $\tilde{\omega}$.
The circular integration contour $\tilde{C}$ with the 
center $\omega_0 = 2\pi c/(1600\, \mathrm{nm})$ and the radius $r_0 = \omega_0\times 10^{-2}$
is used for computing Riesz projections.
(c)~Electric field intensity $|\tilde{\mathbf{E}}|^2$ corresponding to the high-$Q$ resonance.
(d)
Convergence of the eigenfrequency sensitivities $\partial \tilde{\omega}/\partial p_i$
with respect to the polynomial degree $d$ of the FEM ansatz functions.
The sensitivities are computed at the parameter reference values given in Fig.~\ref{fig02}(a).
Relative errors $err_{\mathrm{real},i} =
\Bigl|\mathrm{Re}\left(\frac{\partial \tilde{\omega}}{\partial p_i} (d) - \frac{\partial \tilde{\omega}}{\partial p_i} (d_\mathrm{ref})\right)/
\mathrm{Re}\left(\frac{\partial \tilde{\omega}}{\partial p_i} (d_\mathrm{ref})\right)\Bigr|$, where $d_\mathrm{ref} = 6$.
(e) Relative errors $err_{\mathrm{imag},i}$ for the imaginary parts of the sensitivities; cf.~(d).
}
\end{figure*}

For the calculation of the contour integrals, a numerical integration
with a circular integration contour and a trapezoidal rule is used, which
leads to an exponential convergence behavior with respect to the integration points~\cite{Trefethen_SIAM_Trapz_2014}.
At each integration point, we calculate
$\mathbf{E}$ and $\partial \mathbf{E}/ \partial p_i$ by solving Eq.~\eqref{maxwell}
with oblique incident plane waves as source terms.
The linear functional $\mathcal{L}(\mathbf{E})$ corresponds to a spatial point evaluation of
one component of the electric field, which can be understood as physical observable.
Note that, with Eq.~\eqref{eq:eigenfrequency} and Eq.~\eqref{eq:eig_derivative},
an eigenfrequency $\tilde{\omega}$ and its sensitivity $\partial \tilde{\omega}/ \partial p_i$
can be calculated without solving resonance problems
$\nabla \times \mu^{-1} \nabla  \times  \tilde{\mathbf{E}} - \tilde{\omega}^2\epsilon \tilde{\mathbf{E}}  = 0$
directly. Instead, scattering problems, where Eq.~\eqref{eq:scatt_derivatives}
can be exploited, are solved.
We call the described approach, which combines
Riesz projections and direct differentiation (DD), the {\it Riesz projection DD method}.
Equation~\eqref{eq:eig_derivative} and its numerical implementation exploiting Eq.~\eqref{eq:scatt_derivatives}
are the main results of this work and represent the difference from previous
works on Riesz projections; cf. Ref.~\cite{Zschiedrich_PRA_2018}.

Note that the Riesz projection DD method is not limited to the field of nanophotonics, 
but can be applied to other eigenproblems as well. Maxwell's equations can be replaced by 
another partial differential equation, and then instead of the analytical continuation 
of the electric field $\mathbf{E}$, the analytical continuation of another quantity is evaluated for the contour integration.

\section{Application}
\subsection{Eigenfrequency sensitivities of a nanophotonic resonator}
We investigate an example from the literature, a dielectric nanoresonator 
of cylindrical shape placed on a three-layer substrate, where
constructive and destructive eigenmode interference has been
used to engineer a bound state in the continuum (BIC)~\cite{Koshelev_Science_2020}.
The nanoresonator has been designed taking into account various parameters
to suppress radiation losses:
The radius, the layer thicknesses, and the layer materials
have been chosen to obtain a high-$Q$ resonance.
The nanoresonator is made of the high-index material
aluminum gallium arsenide \mbox{(AlGaAs)} with $20\,\%$ aluminum.
A silicon dioxide ($\mathrm{SiO}_2$) spacer is placed between the nanoresonator
and a film of indium tin oxide (ITO) on a $\mathrm{SiO}_2$ substrate.
A sketch of the designed system is shown in Fig.~\ref{fig02}(a).
For this specific configuration, 
a high-$Q$ resonance with a $Q$-factor of $Q=188 \pm 5$ has been experimentally observed,
and numerical simulations have resulted in $Q=197$,
where the real part of the resonance wavelength
is in the telecommunication wavelength regime,
close to $1600\,\mathrm{nm}$.
The nanophotonic resonator has been exploited as
a nanoantenna for nonlinear nanophotonics~\cite{Koshelev_Science_2020}.

In the following simulations, we consider the constant relative permittivities
$\epsilon_\mathrm{r} = 10.81$ and $\epsilon_\mathrm{r} = 2.084$
for \mbox{AlGaAs} and for $\mathrm{SiO}_2$, respectively, which are extracted from
experimental data~\cite{Koshelev_Science_2020,Malitson_1965}. For the ITO layer,
the Drude model
$\epsilon_\mathrm{r}(\omega_0) = \epsilon_\mathrm{inf} - \omega^2_\mathrm{p}/(\omega_0^2+i \omega_0 \gamma)$
is chosen, where
$\epsilon_\mathrm{inf} = 3.8813$, $\omega_\mathrm{p} = 3.0305\times 10^{15}\,\mathrm{s}^{-1}$,
and $\gamma = 1.2781\times 10^{14}\,\mathrm{s}^{-1}$.
This Drude model is obtained by a rational fit~\cite{Sehmi_2017}
to experimental data~\cite{Koshelev_Science_2020}
and describes the material dispersion of the system.
We further exploit the rotational symmetry of the geometry.
On the one hand, this reduces the computational effort and, on the other hand,
the eigenmodes can be easily distinguished by their azimuthal quantum numbers $m$,
which correspond to the number of oscillations in the radial and axial directions.
When the light sources used for computing Riesz projections
are not rotationally symmetric, such as oblique incident plane waves, the source fields can
be expanded into Fourier modes in the angular direction.
Considering Fourier modes with certain quantum numbers,
only the eigenmodes, eigenfrequencies, and corresponding sensitivities associated
with these quantum numbers are accessed.

\begin{table}[]
	\begin{tabularx}{0.465\textwidth}{crr} \mytoprule
		\hspace{0.2cm}$i$\hspace{0.2cm}
		&\hspace{0.4cm} $\mathrm{Re}(\partial \tilde{\omega}/\partial p_i)\times10^{-10}$ \hspace{0.8cm}
		& $\mathrm{Im}(\partial \tilde{\omega}/\partial p_i)\times10^{-10}$\\
		\mymidrule
		$1$ & $-128.750\,(\mathrm{s\,nm})^{-1} \hspace{1.3cm}$ & $-0.324\,(\mathrm{s\,nm})^{-1}\hspace{0.4cm}$ \\
		$2$ & $-84.568\,(\mathrm{s\,nm})^{-1}\hspace{1.3cm}$ &  $2.660\,(\mathrm{s\,nm})^{-1}\hspace{0.4cm}$ \\
		$3$ & $-7.192\,(\mathrm{s\,nm})^{-1}\hspace{1.3cm}$ &  $-1.955\,(\mathrm{s\,nm})^{-1}\hspace{0.4cm}$ \\
		$4$ & $-0.065\,(\mathrm{s\,nm})^{-1}\hspace{1.3cm}$ &  $0.208\,(\mathrm{s\,nm})^{-1}\hspace{0.4cm}$ \\
		$5$ & $15.047\,(\mathrm{s\,deg})^{-1}\hspace{1.3cm}$ &  $0.039\,(\mathrm{s\,deg})^{-1}\hspace{0.4cm}$ \\
		\mybottomrule
	\end{tabularx}
	\caption{Computed eigenfrequency sensitivities.
	    The sensitivities $\partial \tilde{\omega}/\partial p_i$ 
		correspond to the high-$Q$ resonance of the nanoresonator shown in Fig.~\ref{fig02}(a)
		and are computed at the shown parameter reference values.
		}
	\label{tab:table1}
\end{table} 

We start with computing a Riesz projection to
obtain the eigenfrequency $\tilde{\omega}$ of the high-$Q$ resonance.
Figure~\ref{fig02}(b) shows the complex frequency plane with the calculated
eigenfrequency, $\tilde{\omega} = (1.17309 - 0.00296 i)\times 10^{15}\,\mathrm{s}^{-1}$,
and the corresponding circular integration contour $\tilde{C}$
for the computation of the Riesz projection.
The center and the radius of the contour are selected based
on a-priori knowledge from Ref.~\cite{Koshelev_Science_2020}.
Alternatively, without a-priori knowledge, a larger integration
contour can be used~\cite{Betz_2021}.
The simulations are performed using eight integration points
on the contour $\tilde{C}$, where a sufficient accuracy with respect to the integration points is ensured.
The computations are based on a FEM mesh consisting of $306$ triangles.
To compare the size of the contour with the distances between the eigenfrequencies
within the spectrum of the nanoresonator,
the two eigenfrequencies which are closest to $\tilde{\omega}$ are also shown.
We obtain a $Q$-factor of $Q=198$ for the high-$Q$ resonance, which is in good agreement with the
experimental and numerical results from Ref.~\cite{Koshelev_Science_2020}.
The corresponding electric field intensity $|\tilde{\mathbf{E}}|^2$ is shown in Fig.~\ref{fig02}(c).
The eigenmode $\tilde{\mathbf{E}}$ has the quantum number $m=0$ and 
is strongly localized in the vicinity of the nanoresonator.

Next, the eigenfrequency sensitivities
$\partial \tilde{\omega} / \partial p_i$ with respect to the
parameters $p_1,p_2,\dots,p_5$ sketched in Fig.~\ref{fig02}(a) are computed.
In order to validate the approach,
a convergence study for the polynomial degree $d$ of the FEM ansatz functions
is performed.
Figures~\ref{fig02}(d,e) show the relative errors for the real and imaginary parts, respectively.
Exponential convergence can be observed for all sensitivities with increasing $d$.
The computed sensitivities for $d=5$ are shown in Tab.~\ref{tab:table1}.
Exemplary source code for the Riesz projection DD method and simulation
results are presented in Ref.~\cite{Binkowski_SourceCode_CommPhys}.

\begin{figure}[]
\includegraphics[width=0.485\textwidth]{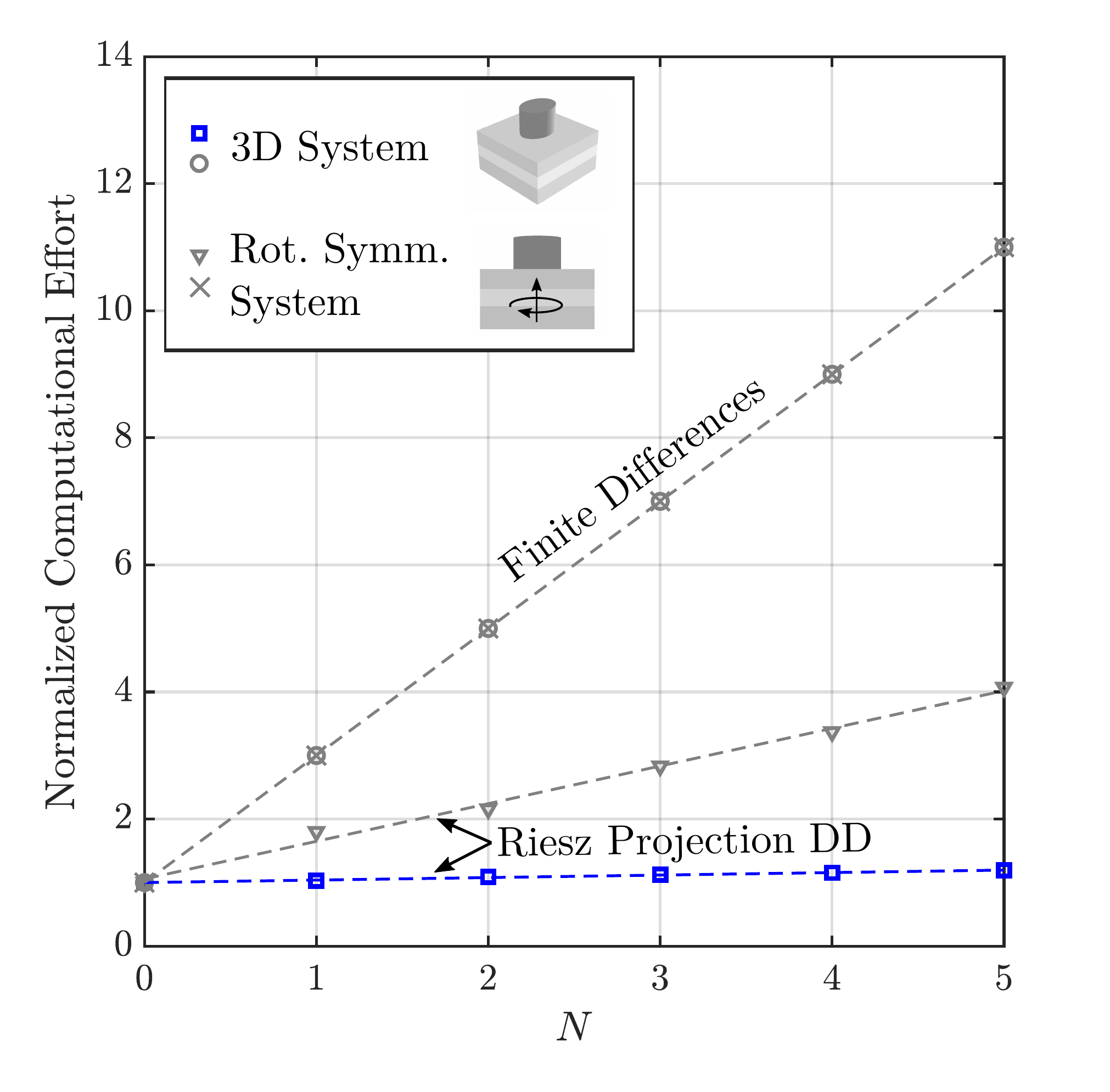}
\caption{\label{fig03} 
Performance of the Riesz projection DD method.
The normalized computational effort over the number $N$ of computed sensitivities
$\partial \tilde{\omega}/\partial p_i$
with respect to parameters $p_1,p_2,\dots,p_N$ is shown.
The sensitivities are computed
at the reference values shown in Fig.~\ref{fig02}(a).
The computational effort is the total CPU time normalized to the 
CPU time spent for computing the eigenfrequency $\tilde{\omega}$,
which corresponds to $N=0$. The time is measured with \textsc{JCMsuite} using four threads on a machine
with a 24-core Intel Xeon Processor running at $3.3\,\mathrm{GHz}$.
For all calculations, to ensure high accuracies, eight integration points at
the integration contour $\tilde{C}$ depicted in Fig.~\ref{fig02}(b) are used.
The degree of the FEM ansatz functions is fixed with $d=5$.
The mesh of the three-dimensional system consists of $4160$ prisms
and the mesh of the rotational symmetric system consists of $306$ triangles.}
\vspace{-0.2cm}
\end{figure}

\subsection{Performance benchmark}
The computational effort of the numerical realization
of the Riesz projection DD method is compared with the computational effort
of the finite difference method. We choose the central difference scheme
$\partial \tilde{\omega} / \partial p_i \approx \left(\tilde{\omega}(p_i+\delta p_i)
- \tilde{\omega} (p_i-\delta p_i)\right)/\left(2\delta p_i\right)$ for the comparison.
Computing central differences is more computationally 
expensive than computing forward or backward differences. However,
more accurate results can be achieved as the error decreases with $(\delta p_i)^2$.
To achieve an adequate accuracy,
sufficiently small step sizes $\delta p_i$ are selected. For example,
for the radius of the nanoresonator,
we choose $\delta p_1 = 0.1\,\mathrm{nm}$.
Note that, also for the finite difference method, we compute
the eigenfrequencies by using the contour-integral-based formula in Eq.~\eqref{eq:eigenfrequency}.

\begin{figure}[]
\includegraphics[width=0.485\textwidth]{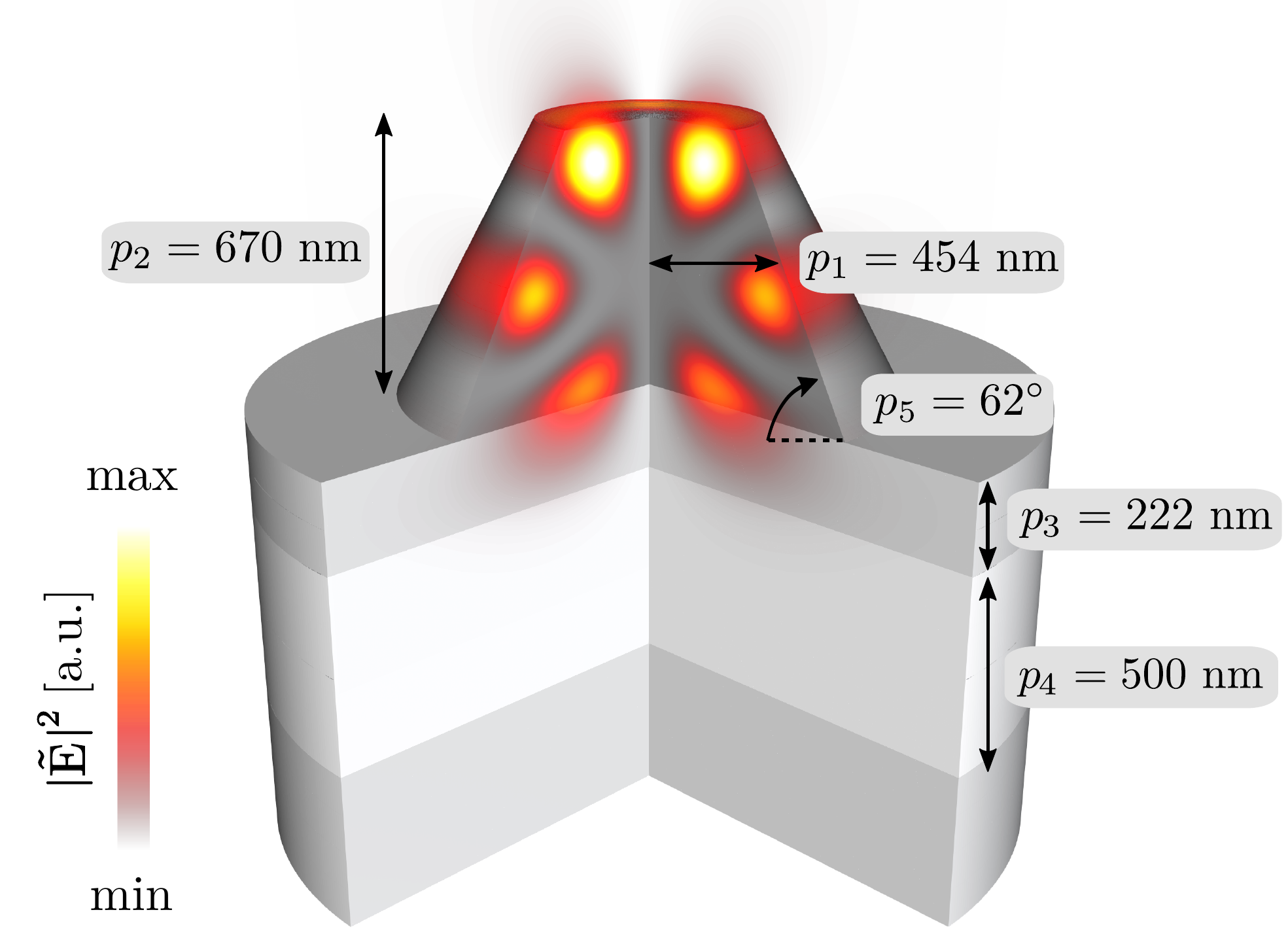}
\caption{\label{fig04} 
Optimization of a nanophotonic resonator.
The optimized nanophotonic resonator with a sketch of 
the electric field intensity $|\tilde{\mathbf{E}}|^2$ corresponding to the high-$Q$ resonance is shown.
The high-$Q$ resonance has a $Q$-factor of $Q=292$. 
The materials of the nanoresonator are the same as for the reference structure in Fig~\ref{fig02}(a).
}
\vspace{-0.2cm}
\end{figure}

We increase the degrees of freedom of the system shown in
Fig~\ref{fig02}(a) by
deforming the cylindrical nanoresonator to an ellipsoidal nanoresonator.
This breaks the rotational symmetry yielding a full three-dimensional system
with new parameters,
the radius of the nanoresonator in $x$ direction and the radius in $y$ direction.
Figure~\ref{fig03} shows,
for the three-dimensional implementation and for
the rotational symmetric implementation,
the normalized computational effort for different numbers of 
computed sensitivities.
We compute the eigenfrequency $\tilde{\omega}$ and
then we add the sensitivities, starting with $\partial \tilde{\omega}/\partial p_1$,
one after the other.
It can be observed that the Riesz projection DD method requires less 
computational effort than the finite difference method,
for any number of computed sensitivities, i.e., for all $N \geq 1$.
In the case of using finite differences,
the computational effort has a slope of about $200\,\%$ because 
for each sensitivity two additional problems with typically the same 
dimension as the unperturbed problem have to be solved.
In the three-dimensional case, a linear regression for the computational effort
gives a slope of about $4\,\%$ for the Riesz 
projection DD method.
The computational effort needed for
the $LU$-decomposition is significant compared to the
matrix assembly and to the other solution steps,
so the possibility of exploiting Eq.~\eqref{eq:scatt_derivatives}
gives a great benefit for the Riesz projection DD method.
For $N=5$, the CPU time required to solve the linear system of equations, 
which includes the $LU$-decomposition, takes $81\,\%$ of the accumulated CPU time.
In the rotational symmetric case, the time for
solving the linear system is negligible.
However, the trend is the same for the three-dimensional and for the computationally cheaper
rotational symmetric case:
The advantage of using Riesz projections significantly increases
with an increasing number of computed sensitivities.

Note that contour integral methods are well suited
for parallelization because the scattering problems can be solved in parallel
on the integration contour. However, as total CPU times are considered for Fig.~\ref{fig03},
this is not reflected by the time measurements.

\subsection{$Q$-factor optimization}
The Riesz projection DD method is applied to further optimize the $Q$-factor of the high-$Q$ resonance of the 
nanophotonic resonator from Ref.~\cite{Koshelev_Science_2020} shown in Fig.~\ref{fig02}(a).
A rotational symmetric nanoresonator is considered
because simulations show that an ellipsoidal shape does not lead to a significant increase of the $Q$-factor.
We use a Bayesian optimization algorithm~\cite{Pelikan_1999}
with the incorporation of sensitivity information.
This global optimization algorithm is well
suited for problems with computationally expensive objective functions and
benchmarks show that providing sensitivities can
significantly reduce computational effort~\cite{Schneider_Benchmark_2019}.
However, other optimization approaches could be used as well.

For the optimization,
we choose the parameter ranges $435\,\mathrm{nm} \leq p_1 \leq 495\,\mathrm{nm}$,
$575\,\mathrm{nm} \leq p_2 \leq 695\,\mathrm{nm}$,
$150\,\mathrm{nm} \leq p_3 \leq 550\,\mathrm{nm}$,
$100\,\mathrm{nm} \leq p_4 \leq 500\,\mathrm{nm}$, and
$60^\circ \leq p_5 \leq 90^\circ$.
To ensure that the optimized nanoresonator
can also be used as nanoantenna in the telecommunication wavelength regime, like the original system,
we add the constraint that the optimized eigenfrequency must lie in the circular contour
with the center $\omega_0 = 2\pi c/(1600\, \mathrm{nm})$ and the radius $r_0 = 4\times 10^{13}\,\mathrm{s}^{-1}$.
In each optimization step,
the Riesz projection DD method is used to compute the eigenfrequency with a quantum
number of $m=0$ lying inside the contour and to calculate the corresponding sensitivities.

A nanoresonator with a $Q$-factor of $Q=292$ is obtained after $61$ iterations of
the optimizer yielding an increase of about $47.5\,\%$ over the original resonator.
More iterations yield only a negligible increase of the $Q$-factor.
The optimized nanoresonator with a sketch of the electric field intensity of its
high-$Q$ resonance and the values for all underlying parameters are shown in Fig.~\ref{fig04}.
The corresponding eigenfrequency
is given by $\tilde{\omega}_\mathrm{opt} = (1.176897 - 0.002015 i)\times 10^{15}\,\mathrm{s}^{-1}$.
Note that, in the optimization domain, the average sensitivity of the $Q$-factor with respect to
the ITO layer thickness $p_4$ is negligible.

\section{Conclusions}
An approach for computing eigenfrequency sensitivities of
resonance problems was presented.
The numerical realization of the Riesz projection DD method relies on
computing scattering solutions and their sensitivities by
solving Maxwell's equations with a source term, i.e., solving linear systems of equations.
This enables direct differentiation for the efficient calculation of eigenfrequency sensitivities.
Although sensitivities of resonances are computed, no eigenproblems have to be solved directly.
The performance of the approach was demonstrated by a comparison with the finite difference method.
The Riesz projection DD method was incorporated into a gradient-based optimization algorithm
to maximize the $Q$-factor of a nanophotonic resonator.

The savings in computational effort are particularly significant for optimization
with respect to several parameters, which is a common task in nanophotonics.
Therefore, we expect the approach to prove especially
useful when many sensitivities are to be calculated.
The Riesz projection DD method can not only be applied to problems in nanophotonics,
but to any resonance problem.

\section*{Data and code availability}
All relevant data generated or analysed during this study are included in this work. 
Tabulated data files 
and source code for performing the numerical experiments can be found in Ref.~\cite{Binkowski_SourceCode_CommPhys}.

\section*{Acknowledgments}
We acknowledge funding
by the Deutsche Forschungsgemeinschaft (DFG, German Research Foundation) 
under Germany's Excellence Strategy - The Berlin Mathematics Research
Center MATH+ (EXC-2046/1, project ID: 390685689)
and the German Federal Ministry of Education and Research
(BMBF Forschungscampus MODAL, project 05M20ZBM).
This project has received funding from the EMPIR programme co-financed by 
the Participating States and from the European Union’s Horizon 2020 
research and innovation programme 
(project 20FUN02 POLIGHT). 
We further thank \mbox{Kirill} \mbox{Koshelev} for providing the experimental
material data for the physical system investigated in this work.

\end{document}